# Confidentiality & Authentication Mechanism for Biometric Information Transmitted over Low Bandwidth & Unreliable channel

*Raju Singh[1]*

(*rajukushwaha36@gmail.com*)

*+91-7503692236*

[1]*School of Computer Engineering and IT, Shobhit University, Meerut, India.*

*A.K.Vatsa[2]*

(*avimanyou@rediffmail.com*)

*+91-9412024636*

[2]*School of Computer Engineering and IT, Shobhit University, Meerut, India.*

*ABSTRACT*

*The security of bio-metric information – finger print, retina mapping, DNA mapping and some other chemical and biological modified genes related information - transfer through low bandwidth and unreliable or covert channel is challenging task. Therefore, Security of biometric information is essential requirement in this fast developing communication world. Thus, in this paper, we propose efficient and effective mechanism for confidentiality and authentication for biometric information transmitted by using arithmetic encoding representation over low bandwidth and unreliable channel. It enhances the speed of encryption, decryption and authentication process. It uses arithmetic encoding scheme and public key cryptography e.g. modified version of RSA algorithm called RSA-2 algorithm.*

Introduction:

**1.1. Biometric Information:** Biometrics, expressed as the science of identifying an individual on the basis of physiological or behavioral traits, seems to achieve acceptance as a suitable method for obtaining an individual's identity [19]. Some of the biometrics information is Finger Print, Iris, Palm Print, Hand Signature stroke etc. With the help of these information we can easily identified the fake person and authorized person. This information is related to human body and their structure which are unique in this earth.[18,19] A fingerprint is made of a series of ridges and furrows on the surface of the finger. The uniqueness of a fingerprint can be determined by the pattern of ridges and furrows. Minutiae points are local ridge characteristics that occur at either a ridge bifurcation or a ridge ending. A ridge termination is the point where a ridge ends sharply. A ridge bifurcation is defined as the point where a ridge forks or diverges into branch ridges.

**1.2. Low bandwidth and Covert Channel:** The bandwidth denotes the number of bits transmitted over the channel in one second. The low bandwidth denotes the very less number of bits transmitted over the channel in one second. The intruder can analyze the traffic easily of covert channel and guess the sender information, due to less number of bits transmitted over the channel. So, this channel cannot provide the security over the important data transmitted over them.

**1.3. Confidentiality:** It is the power to control the secrecy of information. It includes functions to prevent leakage as well as way of secret writing of information. [1], It provides the security to the sender that he can transmit the information over the Network. Any intruder can see the information but it cannot know the meaning of the information except the receiver. [2]The receiver can receive and understand this information by some mechanism with the help secret key or some functions.

**1.3.1 Symmetric Encryption Technique:** It is technique of cryptography by which a sender and receiver can uses a same shared key for the encryption and decryption of information[2]. The sender can encrypted the message and send to the receiver, receiver can uses the same shared key and decrypt the cipher text into plain text. Such encryption techniques are DES, Hill cipher, IDEA etc.

**1.3.2 Asymmetric Encryption Technique:** It is cryptography technique by which a sender and receiver can uses the two different key called public key and private key.[1,2,3] Public key is used for encryption and private key is used for decryption of plain text. The sender encrypt the plain Text using the public key of the receiver and send this cipher text to the receiver, receiver receive this cipher text and decrypt the cipher text using own private key and get back plain text. They use different keys. Such encryption techniques are RSA, Elliptical curve cryptography, etc.

**1.4. Authentication:** [19] Reliable authorization and authentication has become an integral part of our life for a number of daily life applications. Over the internet there are number of unauthorized user to access the secure information transmitted over them. In [5, 12], The sender can provide a security over this information only an authorized user can access this information no intruder or unauthorized user can hack this information. Sender and receiver can uses a same shared ID and password or key which can produce a message authentication code to authenticate the user (sender and receiver).

**1.5. Arithmetic Encoding Technique:** The Huffman and Shannon algorithm produces the integral length code. In paper [13, 14], this algorithm is not required to represent a symbol with a fixed integral code. This algorithm takes the stream of symbol and produces a floating point number this number is less than one and it represent the coded form of plain text. The arithmetic coding has to be an efficient coding technique for compression and communication. [15]This algorithm provides more security, due to probability of character or decimal number is assigned by the user and encoding of data is depending on that probability. So, this method performs better than extended Huffman codes.

**1.6 Problem Identification:** When the biometric information transmitted over the low bandwidth channel there is more chance of information hacked by the intruders. There is no security mechanism is provided by the sender to receive the information is receive by the receiver is not manipulated. The biometric information is most important information for human and some important operations such as military, new research, other security purpose. The low bandwidth channel is most unsecured channel where number of unauthorized users wants to access information. This problem is identified and solved through this paper. This paper provide solution of identified problems by authentication and confidentiality for the bio-metric data transmitted over the low bandwidth or covert channel with the enhancement of speed of encryption and decryption of plain text with RSA-2 Algorithm [1] and authentication code append with cipher text.

**1.7 Paper Organization:** We have organized this paper in Chapters. Firstly we describe the introduction of Bio-metric information, low bandwidth channel, confidentiality, encoding techniques, and authentication etc under the heads of Introduction in Chapter – 1. Subsequently we have gone through the literature review and found problems and solutions in several papers. All this we have mentioned in Chapter – 2 under heads of related works. In chapter – 3 we have discussed our proposed work. Finally we concluded and mention future scope of this paper under heads of Chapter – 4 and Chapter – 5 respectively. We have written all used references during writing of this paper in Chapter -6.

**2. Related Work:** There are many papers which can provide the security, confidentiality and authentication mechanism for the data transmitted over low bandwidth and unreliable channel.

The bio-metric information are most valuable information for human beings, it is expressed as the science of identifying an individual on the basis of physiological or behavioral traits, seems to achieve acceptance as a suitable method for obtaining an individual's identity [19].

In [13], arithmetic encoding technique is used with DES to encrypt the image and transmitted over the covert channel. The arithmetic encoding gives coded data values in between interval of 0 and 1. That gives security and compression over the input files. [13,15]The Arithmetic Coding is extremely efficient, for providing both security and compression simultaneously is growing more important and is given the increasing ubiquity of compressed Bio-metric files in host applications of Defense, Internet and the common desire to provide security in association with these files.

In [2,3] the RSA algorithm is used with some modifications which enhance the speed of RSA algorithm is called RSA-1 and the algorithm which provide security more than RSA algorithm is called RSA-2 algorithm which can enhance confidentiality to the sender. The problem of RSA algorithm is solved [2] through RSA-2 algorithm, it used the numbers instead of character in the plain text are represented by encoding scheme which can be able to represent special character. In case of character and number the intruder can easily know the cipher text and author can replaced it by the special symbols with the help of decimal value into their respective ASCII code character. The RSA-2 algorithm increased the speed of encryption and decryption with enhancement of security also due to special symbol.

In [7, 8], The Diffie-Hellmann key exchange was the first protocol to utilize public key cryptography. The Diffie-Hellmann protocol is used to exchange a secret key between two users over an insecure channel without any previous information between them. The image is transferred with steganography technique and key is used for hiding image information is deliver to receiver with the help of Diffie-Hellmann exchange protocol [7]. The key used in RSA-2 algorithm is delivered to receiver with the help of Diffie-Hellmann algorithm [8-9].

[20] first proposed a remote password-based authentication scheme that could authenticate remote users over an insecure channel. A lot of research has been carried out in the field of Authentication and Key Exchange protocols, which are based on passwords [17, 18]. The Password based user authentication systems are low cost and easy to use .The user chosen passwords are inherently weak since most users choose short and easy to remember passwords. In paper [7], image is transferred to the receiver by Steganography technique with secure key distribution technique.

[5,6], it can provide a more secure authentication technique for sender and receiver with the help of ID and password mechanism. The user of system can register and gets an ID and Password. When any user send data it can encrypt data generate authentication code using receiver ID and password. These codes are appending with cipher text and send this message over the insecure channel.

[5]The Authentication process consist of three phases i) Registration phase ii) Generation of authentication code (sender) iii) verification of authentication code (receiver). The fundamental goals of key exchange are authentication and confidentiality.

[11]Entity authentication inherently depends on some pre-established piece of trusted information; the most common examples include a shared key, a shared password, or a certified public key. [15]Identity-based encryption allows a sender to encrypt a message for a recipient based solely on the recipient's identity. The identities used in identity-based cryptography may be simple usernames, but they could contain more structured information as well.

# 3. Proposed Work:

**3.1 Architecture of Proposed work:** We proposed a security of Bio-metric information using confidentiality and authentication mechanism when we transmit data over the low bandwidth and unreliable channel or covert channel.

**3.2. The working principles for Sender Prospects:** We have transfer the Bio-metric Information over the covert channel, In first step input the Bio-metric information from the User, store it in system and Register the user who uses this service and provide a unique ID and Password for authentication of users. In second step convert this Bio-metric information into decimal number format. These decimal numbers are compressed (encode) with help of arithmetic encoding scheme. Third step take compressed data as input and encrypted with the help of RSA-2 algorithm. Fourth step generates the key for RSA-2 algorithm and Diffie-Hellmann Key Exchange mechanism. The Key used in both the mechanism is same. Fifth steps generate authentication code with the help of receiver ID and password. Sixth steps takes cipher text from encryption and appends it with message authentication code generated in sixth step. After that, this block send this append message over the covert channel. The receiver receives this message.

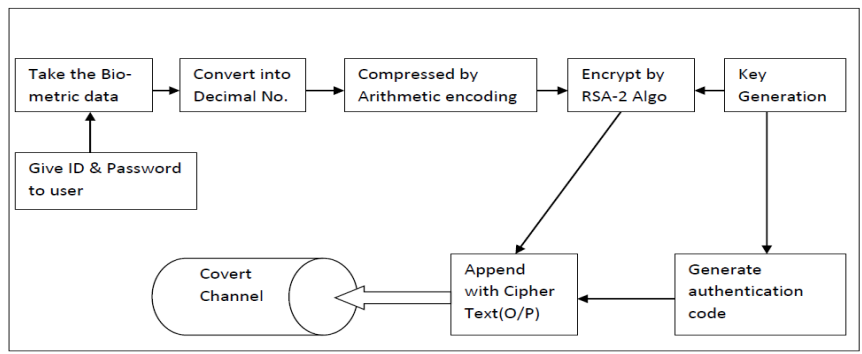

**Figure No 1: Sender Prospects.**

**3.3. The working principles for receiver prospects:** The Cipher Text is receiving at receiver node. He detached the message authentication code form the cipher text and calculates own message authentication code with the help of own ID and password provided at time of registration and compared it with received cipher text authentication code, if it is same then he receive the cipher text otherwise discard it. In, second step accept this cipher text and decrypt with RSA-2 algorithm. After decryption take this result and apply arithmetic decoding technique and decode. Receive key from the sender with help of Diffie-Hellmann Key Exchange algorithm and generate the Key for decryption of cipher text. The key is generated send to the RSA-2 algorithm. In fourth steps take the decimal number result from arithmetic decoding and converted into Bio-metric information. After that original Bio-metric information is obtained at receiver side.

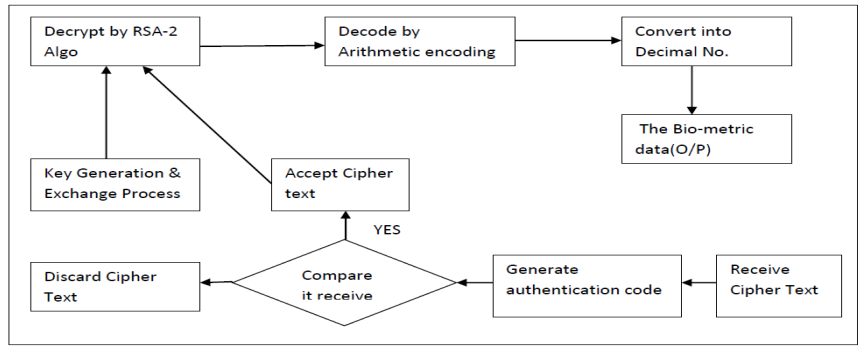

**Figure No 2: Receiver Prospects.**

The steps of the proposed algorithm are as follows:

**Phase – I: Steps for Bio-metrics information:**
- ❖ Take the biometric information as input.
- ❖ Provide a user id and password for the user.
- ❖ Convert into (Decimal)

**Phase - II: Steps for Arithmetic coding:**
- ❖ Set the probability of the Decimal number.
- ❖ Input the Converted form of Bio-metric information.
- ❖ Use arithmetic algorithm, code the input text
- ❖ Set the probability of each alphabet.
- ❖ Take the alphabet from the input text.
- ❖ Initialize all the value
- ❖ Low_value=0.0;
  High_value=1.0;
  Difference=1.0;
- ❖ For every symbol in input repeat
  {
  Low_value =Low_ value+Difference*Range from;
  High_value = High_value+Difference*Range to;
  Difference is = High_value- Low_value;
  }
- ❖ Display Result (Low_value).
- ❖ Take the result of arithmetic coding
- ❖ Store this result

**Phase – III: Step –I: Key generation process:**
- ❖ Select two prime number P,Q;
- ❖ Calculate n=P*Q;;
- ❖ Calculate ϕ(n)= (P-1)*(Q-1);
- ❖ Select integer e; gcd(f(n),e)=1; 1<e<f(n);
- ❖ Calculate d; d=e^-1 mod f(n);
- ❖ Public key KU={e , n};
- ❖ Private key KR={ d , n};

**Step –II: Key Exchange Mechanism:** The sender and receiver can use Diffie-Hellmann key[7,8], Exchange algorithm to exchange key between them. The algorithm is as follows.

**Sender Do this:**
- Choose a prime number P randomly, and choose two integers A and G.
- Compute the public key S= G^A mod P.
- Send this public key S to Receiver.
- Compute the secret value K, as K= R^A mod P.

**Receivers do this:**
- Choose an integer number B randomly
- Compute the public key R= G^B mod P.
- Send this public key S to Sender.
  Compute the Secret value K as K=S^B mod P.

**Phase – IV: Steps for RSA-2 Algorithm.**

This algorithm is focus on to improve security. This algorithm includes the following steps

b) **Encryption**
   Plain text    : M< n
   Cipher Text   : C=M^e (mod n);

c) **Decryption**
   Cipher Text   : C
   Plain Text    : M = C^d (mod n);

**Phase – V: Authentication Mechanism:** The sender can calculate a authentication code(AC) and Then it is append with cipher text [10]. The receiver can receive this header it can calculate the MAC and Matched with receive AC [11,12]; If it is same then the message is received otherwise it can discard the Message.

ID              : It is given to the User at the time of Registration.
Password        : It is given to the User at the time of Registration.
N               : It is multiplication of prime number used in RSA-2 Algorithm.
Message authentication Function    : MAC = (ID*Password) mod N.

After this the cipher text and Key exchange with the help of Diffie-Hellmann algorithm is append with the Authentication code result and send to receiver.

## 4. Result: Sender Prospects:

Suppose we take a sampled Finger prints and convert into decimal numbers with the help of some algorithm and provides a unique ID and Password to the receiver and sender store this ID and Password of that receiver. The ID=**1345** and password=**4679**. Take bio-metric data file and convert into decimal number. After conversion the number obtained is (**4512**). Encode this data value with arithmetic encoding scheme result is(**.7664**). Remove floating point and convert into real values, encrypt this data value with RSA algorithm and result is(**324**). Generate authentication code with Receiver ID and Password with authentication function resultant code is(**144**). Generate secret key for key exchange mechanism the result is(**130**). Convert all these values into respective ASCII character values and arrange all the data values in proper predefine order set by sender and receiver. This paper arrangement is first authentication code, secret key and cipher text. The result is obtained and sends over the channel.

Input Bio metric Data(Finger Impression) → Convert into Decimal number format **(4512)** → After Applying Arithmetic Encoding**(7664)** → Generate Key (**PU**{23,187}, **PR**{7,187}) → After applying RSA-2 algorithm( **324**) → Generate Authentication Code( **144**) → Apply Key Exchange mechanism(**130**) → Make a header for cipher Text(**144130324**). Convert this code into respective ASCII Code. Sender sends this cipher text over a covert channel.

**Receiver Prospects:** The Receiver has following data: ID=**1345** and password=**4679** and PU {**23,187**}. Receive cipher text and convert into ASCII character to respective decimal values and segment all the data values from order defined by the sender. Generate authentication code from ID and password if it matched with receive authentication code accept cipher text, otherwise discard it. When it accept cipher text decrypt with RSA algorithm and result is(**7664**), convert this real valve into floating point value by adding decimal point at starting of number, decode this data values with arithmetic encoding technique and obtained result is (**4512**),convert this decimal data values into bio-metric information using algorithm.

Cipher Text(**14413057**) ⟶ Remove Header and Key Exchange ⟶ Check the authentication code by ID and password(**144**) ⟶ Cipher Text(**324**) ⟶ Decrypted (**7664**) ⟶ Apply Arithmetic Decoder and decode Plain Text(**4512**) ⟶ Take this Decimal Number(**4512**) Convert into Bio-metric Information.

**5. Conclusion:** In this paper, we are providing a security, confidentiality and authentication over bio-metric information when it transmitted over low bandwidth and unreliable channel. The bio-metric information is converted into integer real values and encodes this value with arithmetic coding. Which produces integral length code and result is in floating point number. This can provide more security because the probability of alphabet or decimal number is set by the user and it is known only to authenticated person. The confidentiality is provided through RSA-2 algorithm, this enhances the speed of the encryption and decryption. The speed of RSA-2 algorithm is depending on size of bio-metric file. This algorithm is highly secure as compared to RSA algorithm because after encryption of data it again encrypted into respective decimal value into ASCII character. The Diffie-Hellmann algorithm is used for secure transmission of key between the sender and receiver. This can deliver the key between the sender and receiver with secure transmission over the low bandwidth channel. The key is also send with cipher text and receiver computes secret key. When sender wants to change key at any instance he change its secret key and send to receiver with proper message add with it. The User ID and password mechanism is used for to authenticate the sender and receiver. Sender generates authentication code and appends with cipher text and sends to receiver. Every time he send message he append this code, this code is also useful for identifying specific receiver due to specific receiver ID and Password. This algorithm is useful for many application of secure data transmission over the low bandwidth channel, such as military, intelligence services and other important operations.

**6. Future Scope:** This paper is used for bio-metric data transmission on unsecured channel, to make more effective and efficient can use another compression algorithm and make some changes in the RSA-2 algorithm which can provide more confidentiality, security and also increase speed of encryption and decryption of plain text. To provide more security over authentication improve the authentication code function and take another authentication system instead of User ID and password, such as bio-metric data for authentication, smart card etc. Here the mechanism is used for transferring the Bio-metric data but other can use it for different other purpose also.

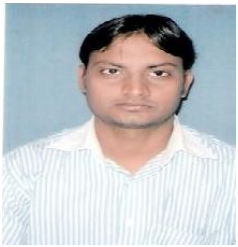
**Raju Singh** is working as a Lecturer in Department of CSE\IT at GNIT Girls Institute of Technology, Greater Noida. U.P. India. He obtained his B.Tech in Computer Science & Engineering from B.S.A. College of Engineering & Technology, Mathura, UP, India and Pursuing M.Tech in Computer Engineering from Shobhit University, Meerut, UP, India. His area of interest is Network Security, Congestion control.

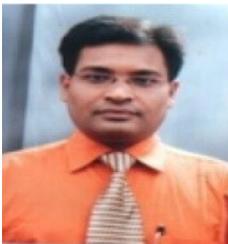
**A.K.Vatsa** is working as Assistant Professor in the School of Computer Engineering and Information Technology at Shobhit University, Meerut (U.P.). He obtained his M-Tech(C.E.) from Shobhit University and B-Tech(I.T.)from V.B.S. Purvanchal University, Jaunpur (U.P.). He has worked as software engineer in software industry. He has been in teaching for the past one decade. He has been member of several academic and administrative bodies. During his teaching he has coordinated several Technical fests and National Conferences at Institute and University Level. He has attended several seminars, workshops and conferences at various levels. His several papers are published in various national and international conferences across India. His area of research includes MANET (Mobile Ad-Hoc network), Congestion Control, Network Security and VOIP-SIP (Voice over IP).